\documentclass{trbunofficial}
\usepackage{graphicx}
\usepackage{gensymb}
\usepackage{graphicx}
\usepackage{hyperref}
\usepackage{listings}
\usepackage{setspace}
\usepackage{titlesec}
\usepackage{times}

\AuthorHeaders{Farabi}
\title{ Heart rate and respiratory rate
prediction from noisy real-world smartphone based on Deep Learning methods}

\begin{document}
\author{%
  \textbf{Ibne Farabi Shihab}\\
  Graduate Student\\
  Department of Computer Science\\
  Iowa State University of Science and Technology\\
  Ames, IA\,50011-1066, USA\\[6pt]
  \texttt{ishihab@iastate.edu}%
}

\maketitle            


\TotalWords{200}

\section{Abstract}

Using mobile phone video of the fingertip as a data source for estimating vital signs such as heart rate (HR) and respiratory rate (RR) during daily life has long been suggested. While existing literature indicates that these estimates are accurate to within several beats or breaths per minute, the data used to draw these conclusions are typically collected in laboratory environments under careful experimental control, and yet the results are assumed to generalize to daily life. In an effort to test it, a team of researchers collected a large dataset of mobile phone video recordings made during daily life and annotated with ground truth HR and RR labels from N=111 participants. They found that traditional algorithm performance on the fingerprint videos is worse than previously reported (7 times and 13 times worse for RR and HR, respectively). Fortunately, recent advancements in deep learning, especially in convolutional neural networks (CNNs), offer a promising solution to improve this performance. This study proposes a new method for estimating HR and RR using a novel 3D deep CNN, demonstrating a reduced error in estimated HR by 68\% and RR by 75\%. These promising results suggest that regressor-based deep learning approaches should be used in estimating HR and RR.

\hfill\break%
\noindent\textit{Keywords}:  Vital signs, deep learning, regression, mo-
bile phones, mHealth, photoplethysmography
\newpage

\section{Introduction}

The tracking of vital signs, such as heart rate (HR) and respiratory rate (RR), has become increasingly prevalent in daily life, serving as a general measure of human health and for quantifying symptoms in specific conditions such as atrial fibrillation \cite{b1}, panic attacks \cite{b2}, chronic obstructive pulmonary disease, asthma \cite{b3}, and post-operative recovery \cite{b4}. Fortunately, with the widespread availability of mobile phones, many individuals now have access to technology for taking these measurements without needing expensive and inconvenient companion devices, thus enabling measurements in resource-limited environments. In particular, the use of mobile phone videos for measuring vital signs has emerged as a promising option due to the near-ubiquity of smartphones, making it a convenient option for the elderly population. This approach eliminates the need for a smart wrist or any other devices that use a PPG signal. Overall, using mobile phone technology to track vital signs during daily life offers great potential for improving health monitoring and disease management. As such, continued research in this area is crucial to ensure the accuracy and reliability of these measurements and extend this technology's reach to those who may benefit most.

 Mobile phone video of the fingertip has emerged as a promising alternative to the photoplethysmogram (PPG) for estimating vital signs such as HR and RR \cite{b5},\cite{b6},\cite{b7},\cite{b8},\cite{b9},\cite{b10}. Previous studies have demonstrated high accuracy (e.g., median error in RR was 0.5\% in \cite{b10} and maximum observed error in HR estimates ranged from 1.8 to 7.5 beats per minute for a variety of methods in \cite{b11}) in estimating these quantities using signal-processing approaches in the frequency and time domains. However, these studies typically employ controlled testing conditions, which do not fully capture the complexity of real-world data collected during daily life.

Recent advances in deep learning have demonstrated significant improvements in various traditional computer vision tasks, including image classification \cite{b12},\cite{b13} and object detection \cite{b14},\cite{b15}. Deep neural network architectures such as the convolutional neural network (CNN) have been successfully applied to regression problems, achieving state-of-the-art results in complex vision tasks such as head and human pose estimation \cite{b16},\cite{b17},\cite{b18} or facial landmark detection \cite{b19}. These deep learning techniques have been shown to learn powerful distributed feature representations for complex datasets without requiring precise human-engineered input representations. Given their success in previous regression problems, CNNs may also be well-suited for estimating HR and RR from mobile phone video, despite the additional complexity of unconstrained measurements (e.g., variable lighting conditions, video quality). However, current deep learning algorithms primarily work in the spatial and temporal domains, and this study is focused on calculating HR and RR from fingerprint videos, which require a more significant emphasis on the temporal domain.  Furthermore, there is a growing interest in applying deep learning techniques to medical science. Despite this, there has been limited advancement in measuring HR and RR, with PPG remaining a commonly used but outdated and sensitive method. With these considerations in mind, this research proposes a unique approach for measuring HR and RR from mobile phone video of the noisy fingerprint, utilizing the power of deep learning. This new method offers the potential for improved accuracy and robustness, enabling accurate measurement of vital signs in a wider range of environments and populations. Overall, this promising advancement highlights the potential for continued innovation in medical science by integrating cutting-edge technology and novel approaches.From a deep learning perspective, this research will also focus on the spatial domain at the initial stage to address the noise issue.

This paper presents a novel deep-learning approach for measuring HR and RR from real-world mobile data, which is often noisy and complex. While 3D deep CNN are typically used for action recognition tasks involving spatial and temporal data, the architecture proposed in this paper is specifically designed to primarily process HR and RR data in the temporal domain. Additionally, this research introduces a new dataset of daily recordings by participants, who tracked their ground truth RR and HR. This study is compared  to a recently developed signal processing-based approach only used with the data collected in controlled clinical settings due to the lack of open-sourced work based on the process followed.

\subsection{LITERATURE REVIEW}
The use of video cameras to estimate HR has been explored in early research, with one study estimating HR by evaluating changes in blood flow to the face using a video camera, leading to further exploration of techniques for extracting physiological information from videos \cite{b27}. Researchers have focused on adopting optical flow analysis to identify skin color changes caused by blood flow, with one study developing an algorithm to track variations in facial color to derive heart rate data \cite{b28}, while another traced color changes on the fingertip due to variations in pulse \cite{b29}. 

Later studies have explored the application of machine learning to video data analysis, with researchers deriving heart rate information from facial videos with 98.5\% accuracy \cite{b30}, and tracking respiration rate from chest-related videos. In medical applications, researchers have developed machine learning systems for monitoring heart and respiration rates in neonatal intensive care settings, achieving high levels of accuracy, which is 95\% \cite{b31}. Taking a step further, some researchers have explored the use of deep learning to predict heart rates.

One study used real-time deep learning and stream processing systems based on long short-term memory (LSTM) to predict heart rate by drawing on electrocardiogram (ECG) numbers, achieving high accuracy in the form of a mean absolute error of one beat per minute in real-time \cite{b32}. Another study developed a deep learning model for real-time heart rate prediction from ECG data using CNN processed with the Apache Flink platform, achieving a high accuracy of 97.4\% \cite{b33}. Additionally, deep learning has been combined with adaptive filtering algorithms to remove noise from ECG data, resulting in an accuracy of 98.3\% \cite{b39}. A wavelet neural network (WNN) based deep learning model and the Spark Streaming stream processing platform have also been used to predict heart rate with a 98.5\% accuracy, proving adept at processing large volumes of real-time ECG data \cite{b33}.

Few studies in this area also utilized the pulse-respiration quotient to predict heart and respiratory rates using video data, resulting in higher accuracy and success in curbing motion artifacts compared to conventional approaches. Several studies have since merged pulse respiratory quotient and deep learning to predict heart and respiratory rates from video data, increasing accuracy and demonstrating the potential for continued innovation in this field \cite{b40}.

Even though all the studies have shown significant results in predicting HR and RR, all the studies have been done in a controlled environment where noise from real data has been ignored totally.

\section{Experimental Platform}

This section will describe the data source, the platform, and the models used in this study.

\subsection{Data Collection}
 In this study, featuring N=105 participants, researchers employed a custom web form to collect video data, instructing subjects to record two 30-second videos—captured before and after a minute of physical activity—with their index finger pressed against an iPhone 6 (or newer) camera lens and the flash enabled. Simultaneously, participants counted their RR  while an expert measured their HR through radial or carotid pulse palpation. 161 videos were ultimately submitted for analysis, as some subjects provided only one video. Measured values during video recording were doubled to obtain beats and breaths per minute. It has been demonstrated that 30-second radial pulse measurements offer efficient HR measurements, with a mean absolute error of slightly over four beats per minute (BPM)\cite{b20}.  Notably, data collection occurred in everyday settings without experimenter oversight, ensuring a more accurate representation of real-world data quality and aligning with the study's goal of estimating HR and RR from noisy, real-life data. Due to the sensitive nature of the data, researchers refrained from open-sourcing it.
\subsection{Data Pre-processing} \label{datapreproc}
Before subjecting each video, along with its respective HR and RR labels, to the neural network for training or testing, a series of pre-processing steps are undertaken. These measures not only eliminate malformed data but also standardize the structure and diminish the complexity of each video clip. The Python OpenCV and Python Image Library (PIL) modules were employed to implement all video pre-processing code.
\subsubsection{Video Pre-processing}  
In the initial phase of video pre-processing, each of the 161 video clips is scrutinized to verify the presence of a "beating" or "pulsing" motion. Employing an automated signal processing method \cite{s4}, 71 clips (44\%) were determined to be of adequate quality. When deployed, this stage could be achieved through a simple binary classifier that distinguishes quality videos from those lacking the desired motion. \\
\indent Following the removal of poor-quality videos, the remaining clips undergo standardization. Videos recorded on newer iPhone models (8 or higher) are converted from 60 to 30 frames per second(using FFMPEG \cite{b21}). Additionally, each video's first and last two seconds are truncated to eliminate artifacts associated with finger repositioning and removal.
\newline \indent The culmination of the video data pre-processing involves breaking down each video into its constituent frames and resizing them to more manageable dimensions. In this instance, frames are down-sampled to $32\times32\times3$. The rationale for using a lower resolution than typical smartphone cameras in HR and RR prediction tasks stems from two factors:

\begin{enumerate}

\item Given the homogeneity of the video data (consisting of varying shades of red), the spatial features of each frame convey limited semantic information about the temporal color variations, which the network relies on for learning.
\item Higher spatial dimensions correspond to increased pixel counts, resulting in greater computational overhead for training and subsequent video processing. As the task primarily focuses on measuring color change frequency, down-sampling each frame aids in reducing computational costs.
\end{enumerate}

Upon completing the pre-processing of video data, the collection of frames is stored on a hard disk, and a catalog file is generated for future reference.

\subsubsection{Label Adjustment}
The labels associated with each valid video clip identified a considerable degree of noise. Upon examination of the video data and labels, many clips were discovered to be marginally longer or shorter than the 30-second mark. Since the HR (beats per minute) and RR (breaths per minute) values were derived by doubling the counts recorded during the video's duration, minor timing discrepancies resulted in errors in the ground truth HR and RR. Consequently, for a video of length t seconds, the adjusted HR $\mathrm{HR}$ and RR $\mathrm{RR}$ values—expressed in beats per minute and breaths per minute, respectively—are calculated as follows:

\begin{align*}
\mathrm{HR} &= \frac{\mathrm{HR}}{t}(60),\ 
\mathrm{RR} = \frac{\mathrm{RR}}{t}(60)
\end{align*}
\subsection{Dataset} \label{dataset}
\begin{figure}
\includegraphics[width=0.7\textwidth]{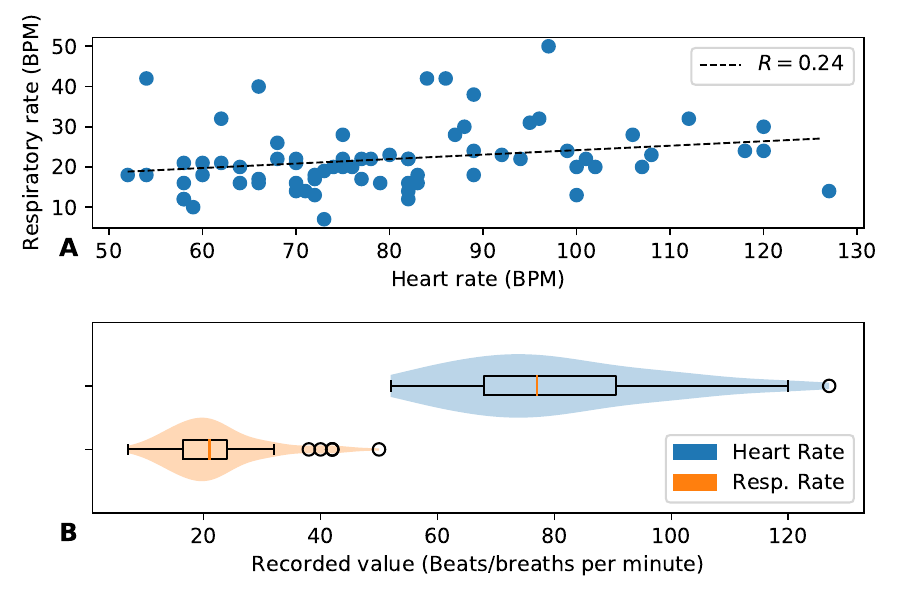}
\caption{ Slight positive correlation exists between the observed HR and RR (\textbf{A}). Observation densities for each are included in the violin plot of (\textbf{B}). Note that RR observations are skewed right with several outliers.}
\label{fig:train}
\end{figure}

\begin{figure}
\includegraphics[width=0.7\textwidth]{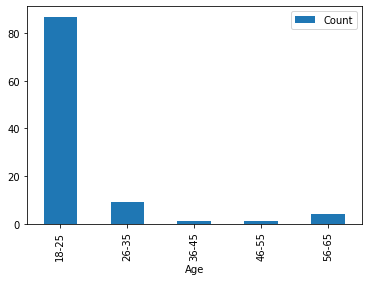}
\caption{Age distribution of subjects}
\label{fig:ageDist}
\end{figure}

The dataset employed for training and testing comprises the quality-tested and pre-processed frames outlined in Section \ref{datapreproc}. Encompassing 71 individual video clips from 46 unique subjects, the dataset features a mean length of 810 frames, amounting to 72,156 separate images or approximately 40 minutes of video at 30 FPS. The recorded HR exhibit a mean $\mu_{\mathrm{HR}}=81$ and standard deviation $\sigma_{\mathrm{HR}}=17.7$ while the RR have mean $\mu_{\mathrm{RR}}=22$ and a standard deviation $\sigma_{\mathrm{RR}}=8.3$. The relation between recorded HR and RR for each observation and the distribution of recorded HR and RR are depicted in Fig. \ref{fig:train}. Additionally, the age of subjects plays a significant role in determining HR and RR; thus, the age distribution of subjects is illustrated in Fig. \ref{fig:ageDist}. 

\section{Methodology}

\subsection{Network Architecture} \label{netarch}

Achieving accurate HR and RR estimation necessitates an architecture capable of capturing both spatial and temporal information within each video. This investigation examined recurrent neural networks (RNN) and CNN, with the CNN-based approach likely outperforming RNN-based methods due to the latter's requirement for extensive training data (e.g., \cite{b22},\cite{b23},\cite{b24},\cite{b25}).\\

Drawing inspiration from the renowned C3D architecture, the researchers developed a lighter-weight version called Deep Video Regression 1 (DVR 1) for HR and RR prediction tasks. This 3D CNN is designed to learn spatiotemporal features for video analysis tasks. The network processes a sequence of 360 frames, each with dimensions of $32\times 32 \times 1$, where the third dimension represents either the red channel or a grayscale composite of the frame. Periodic batch normalization layers are incorporated after each convolution to mitigate vanishing or exploding gradients. Each 3D-convolutional layer is succeeded by a 3D max pooling layer with stride (1, 2, 2) to reduce computational costs and extract low-level features in a small temporal neighborhood. A small stride is crucial for counting purposes. The final convolutional filters pass through three fully connected layers, each featuring a dropout of 0.15 as a regularization parameter and a single output. The ReLU activation function is employed in all layers, except for the final output with linear activation. The outcomes of this network are discussed in Section III ( Table \ref{fig:hrperf1-table}, \ref{fig:rrperf1-table} and \ref{fig:hrrrperf1-table} ).

\begin{figure}
\begin{center}
\begin{tabular}{ c c c}
\hline
 \textbf{Layer} & \textbf{\# Units} & \textbf{Kernel/Pool Size}\\ 
 \texttt{Input} & $360 \times 32 \times 32 \times C$ & - \\
 \texttt{Conv1} & 64 & $(90,5,5)$   \\  
 \texttt{MaxPool1} & - & $(1,2,2)$ \\
 \texttt{BN1} & - & - \\
 \texttt{Conv2} & 64 & $(1,5,5)$   \\ 
 \texttt{Conv3} & 64 & $(60,1,1)$   \\
 \texttt{MaxPool2} & - & $(2,2,2)$ \\
 \texttt{BN2} & - & - \\
 \texttt{Conv4} & 64 & $(1,3,3)$   \\
 \texttt{Conv5} & 64 & $(30,1,1)$   \\
 \texttt{MaxPool3} & - & $(2,2,2)$ \\
 \texttt{BN3} & - & - \\
 \texttt{Conv6} & 128 & $(1,3,3)$   \\
 \texttt{Conv7} & 128 & $(15,1,1)$   \\
 \texttt{MaxPool4} & - & $(1,2,2)$ \\
 \texttt{BN4} & - & - \\
 
 \texttt{Conv8} & 256 & $(1,3,3)$   \\
 \texttt{Conv9} & 256 & $(10,1,1)$   \\
 \texttt{MaxPool5} & - & $(2,2,2)$ \\
 \texttt{BN5} & - & - \\
 
 \texttt{FC1} & 512 & - \\
 \texttt{FC2} & 512 & - \\
 \texttt{FC3} & 256 & - \\
\end{tabular}
\end{center}
\caption{The 3D Deep Video Regression architecture 3(DVR 3) used for the heart rate and respiratory rate prediction tasks.}
\label{fig:architecture}
\end{figure}

\begin{figure}
\includegraphics[width=1\textwidth]{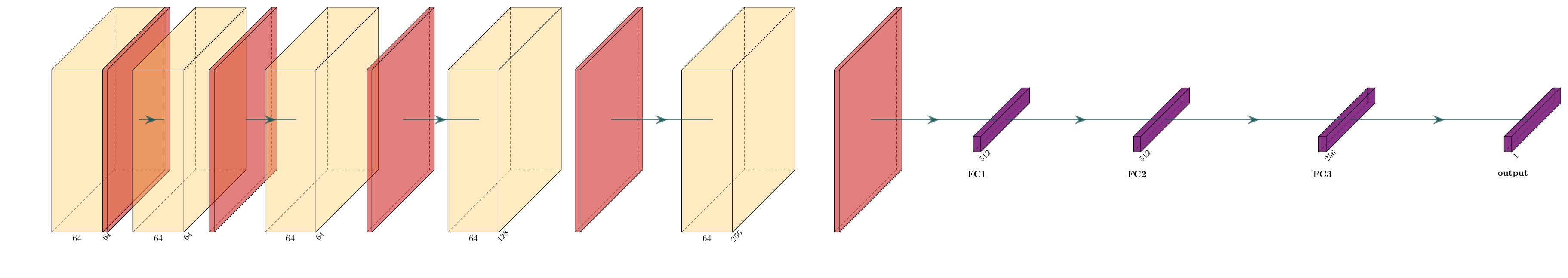}
\caption{Pictorial view of the 3D Deep Video Regression architecture 3(DVR 3) used for the heart rate and respiratory rate prediction tasks}
\label{fig:reg}
\end{figure}

 As it is a counting problem, the network requires sufficient time to accurately count. Moreover, the scales of HR and RR differ, with DVR 1 biased towards HR (see Section III). To address these challenges, this study proposes a different approach (DVR 2) as detailed below.

Initially, 360 frames are fed into the networks in one go. This is because calculating RR in the dataset under consideration is challenging, with a mean of 22 and a standard deviation of 8.3 (see Section II(C)). To accommodate this, 6 seconds of the video (360 frames) are processed at once, ensuring at least 2 bits per second for the respiratory rate. DVR 2 employs a larger depth size compared to DVR 1, focusing on temporal information. A larger 5x5 kernel is also utilized to account for the uncertainty in finger placement. The depth of the convolutional layers is gradually decreased from 90 to 10, allowing the network ample time for counting or reasoning color variation frequency. Additionally, another two convolution layers are implemented before reducing the kernel size, as maintaining the same size throughout the network would be computationally expensive. The first two layers effectively capture finger position. The computational issue arising from the in-depth approach is evident in Table \ref{fig:train}. Apart from these changes, the DVR 2 settings remain the same as DVR 1. The results for DVR 2 are reported in Section III (Table \ref{fig:hrperf1-table}, \ref{fig:rrperf1-table}, and \ref{fig:hrrrperf1-table}).

To address the computational issue in DVR 2, 1x1 filters (excluding the first layer) are introduced in DVR 3. By splitting each convolution layer into two convolution layers with 1x1 filters, the number of trainable parameters is significantly reduced while maintaining the same results as DVR 2 (see Table \ref{fig:train}). DVR 3 also requires less training time compared to DVR 1 and DVR 2 (see Table \ref{fig:train}). For this research, two consecutive convolutional layers are treated as one layer, as they are split for computational purposes.

\begin{table}
\begin{center}
\begin{tabular}{c c c c}
Unit & DVR 1 & DVR 1 & DVR 1\\
\hline
Params & 355.6 M & $ 470.6 M $ & $ 157.5 M$\\
Time & 32 hours &  42 hours & 27 hours 
\end{tabular}
\end{center}
\caption{Trainable parameters and time comparison of  3 networks}
\label{fig:para-table}
\end{table}

Unlike action recognition tasks that demand deeper networks due to the complexity and variance between adjacent video frames, the challenge in this research is not as intricate. Rather, this study requires a distinct approach that delves deeper while maintaining minimal computation, considering the subtle signals given by individual video frames in any given clip from the dataset. The primary objective of our network is to concentrate on learning the frequency of color variation within each video. From this frequency, the network should determine a valid transformation to estimate the HR and RR accurately.

\subsection{Training and Testing Details} \label{train_and_test}
Prior to training, 20\% of the study subjects (9/52 subjects, 15/71 videos) were set aside to test model generalizability. No subject spanned both training and testing sets to minimize the possibility of subject-specific features that could unfairly increase testing accuracy.

The network described in \ref{netarch} is trained using a $4$-fold sorted stratified cross-validation. The value K=4 for cross-validation was chosen to provide an approximately 55-25-20 train-validation-test split for each fold, which was empirically shown to yield the best results. The fold generation process starts by sorting the data $(\mathrm{video}, \mathrm{HR}, \mathrm{RR})$ tuples in descending order by multiplying $\mathrm{HR} \times \mathrm{RR}$. The sorted list of tuples is then divided into 4 approximately equally-sized tiers. All folds are created by initializing each one as an empty list of tuples and sampling uniformly without replacement from each tier $N/K$ samples, where N is the number of videos in the training set.

During training, the network is fed a batch of 5,720-frame sequences. Frame sequences are chosen by first randomly selecting a clip from the training pool and then sampling a 720-frame sequence. Each 720-frame sequence is down-sampled to produce a 360-frame sequence across 24 seconds of video. Down-sampling reduces computational costs and removes redundant information captured in adjacent frames.

Various forms of on-the-fly augmentation were implemented to increase variability in spatiotemporal features and discourage the network from overfitting the training set. As samples were shown to the network, spatiotemporal features were randomly perturbed by applying different types of transformations, including:

\begin{itemize}
\item{\textit{Vertical flip}. With probability 0.5, the entire sequence of frames is flipped on the vertical axis.}
\item{\textit{Horizontal Flip}. Like \textit{Vertical Flip}, with probability 0.5, the entire sequence is flipped horizontally.}
\item{\textit{Rotation}. Rotate the entire sequence by $D$ degrees, where $D$ is a randomly chosen integer from  the interval between 0 and a limit value $L \leq 360$. This research uses a $L=90$\textdegree.}
\item{\textit{Zoom}. Zoom into or out of each frame in the sequence by a randomly chosen percentage between $-z$ and $z$, a chosen parameter. }
\item{\textit{Vertical Shift}. Shift the vertical focus by a specified parameter, a random percentage between $v$ and $-v$. A negative vertical shift moves the focus downward. The pixels that are cropped out by this operation are replaced by either the topmost or final row of pixels in the frame.}
\item{\textit{Horizontal Shift}. Shift the horizontal focus by a random percentage between $h$ and $-h$, a specified parameter. A negative horizontal shift moves the focus to the left. The leftmost or rightmost column of pixels replaces any cropped pixels by this process.}
\item{\textit{Brightness Shift}. Increase the brightness of each frame by a random factor selected from the interval $[b_{1}, b_{2}]$, where each $b_{i} \in [0, 1]$ and $b_{1} < b_{2}$.}
\end{itemize}

 When pixels are cropped during augmentation, they are replaced with the values of their closest valid pixels. Furthermore, all listed augmentation methods are applied to the entire input sequence rather than at the spatial level. Zoom, vertical shift, and horizontal shift parameters are chosen at $z=v=h=50\%$, with a brightness range of $[0.1, 1.0]$.

 For each experiment and across all folds, network weights are initialized using He \cite{s1} initialization and updated with the Lookahead \cite{s2} optimization method. An initial learning rate of $1 \times 10^{-5}$ and a slow step size of 0.5 are used, informed by a mean squared error loss function between the prediction and actual label. Each fold is trained for 40 epochs, with the provision that 10 epochs without a decrease in validation loss will cause early termination of training. Model training and data manipulation code were implemented using the Keras Deep Learning API with a Tensorflow backend. Training occurred on four separate GPU-enabled machines, each equipped with two NVIDIA 3080 GPUs. Lastly, DP-SGD was used as an optimizer for the models to protect them from leaking subject information \cite{s5}. A detailed explanation of this optimizer's use is beyond the scope of this paper but can be found in the cited work.

\section{Results}

The deep 3D convolutional network described in \ref{netarch} is trained to accomplish three different tasks in six different experiments: prediction of HR, RR, or both by taking as input 6 seconds of pre-processed video as described in \ref{datapreproc} and either extracting the red channel from video frame or converting the sequence of frames to grayscale (computed by combining the red (R), green (G), and blue channels (B) as per $GRAY = 0.21R + 0.72G + 0.07B$). 
\newline  In each experiment, the network is trained and tested on the same training and testing sets, with 4 cross-validation folds as detailed in \ref{train_and_test}. Each fold's overall performance is evaluated by computing the mean squared error (MSE) on a validation subset of the training data. At test time, none of these augmentations are applied to the input data. For each experiment, the fold that performs the best on the validation set is taken as the final model and its performance is characterized on the 15 held-out test videos. 
\newline  For comparison, this study also extracted the red channel from each video, averaged the pixels across each frame, and computed estimates of HR and RR using the recently proposed ensemble empirical mode decomposition with principal component analysis (EEMD-PCA) approach presented in \cite{s3} for each video in the testing set. This method was chosen as it has been shown to estimate HR and RR from PPG data more accurately than the existing methods. 
\newline  The performance of each network and the EEMD-PCA method on the withheld test data is determined by computing the root-mean-square (RMS) difference between each method's predicted HR and RR and the ground truth values. This study further examined the performance of the deep networks by constructing Bland-Altman and correlation plots where appropriate. 
\subsection{Heart Rate Prediction}
The network is tasked with outputting a single real-valued number representing the predicted HR in beats per minute of the subject that recorded the video.
\newline  Table \ref{fig:hrperf-table} shows that the top-performing fold from the grayscale training method outperforms the red channeled data by a significant margin in the loss. The root means squared error suggests that the best performing of the two models is approaching human accuracy of estimation on 24 seconds of data. 

\begin{table}
\begin{center}
\begin{tabular}{c c c c}
Channel & CNN MSE & CNN RMS & EEMD RMS\\
\hline
Gray & 33.37 & $ 5.71$ & $ 9.02$\\
Red & 99.67 & $ 9.98$ & $ 9.02$ 
\end{tabular}
\end{center}
\caption{DVR 1 Performance (MSE and RMS) of the HR predictions on test data for gray and red channeled models and the EEMD-PCA method. The model that uses grayscaled data performs significantly better than both its red counterpart and the EEMD-PCA method.}
\label{fig:hrperf1-table}
\end{table}

\begin{figure}
\includegraphics[width=0.8\textwidth]{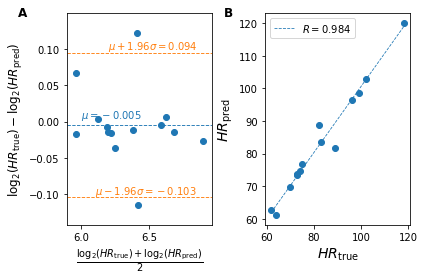}
\caption{Bland-Altman plot (\textbf{A}) illustrates the excellent agreement between the predicted ($\mathrm{HR}_{\mathrm{pred}}$) and ground truth ($\mathrm{HR}_{\mathrm{true}}$) heart rates for the grayscale heart rate model on the test set. Similarly, the correlation plot (\textbf{B}) reflects the excellent agreement between the predicted and actual heart rates.}
\label{fig:hrperf-plot}
\end{figure}

\begin{table}
\begin{center}
\begin{tabular}{c c c c}
Channel & CNN MSE & CNN RMS & EEMD RMS\\
\hline
Gray & 8.53 & $ 2.92$ & $ 9.02$\\
Red & 25.96 & $ 5.10$ & $ 9.02$ 
\end{tabular}
\end{center}
\caption{DVR 2 and 3 Performance (MSE and RMS) of the HR predictions on test data for gray and red channeled models and the EEMD-PCA method. The model that uses grayscaled data performs significantly better than both its red counterpart and the EEMD-PCA method.}
\label{fig:hrperf-table}
\end{table}
The research further investigated the error rate of the grayscale model in Fig. \ref{fig:hrperf-plot} by considering the Bland-Altman plot (agreement between predictions and true values) and correlation (strength of linear relationship) of predictions on the test set. Both high agreement and high correlation show that the network is in fact an accurate predictor of HR.
\subsection{Respiratory Rate Prediction}
\begin{figure}
\includegraphics[width=0.7\textwidth]{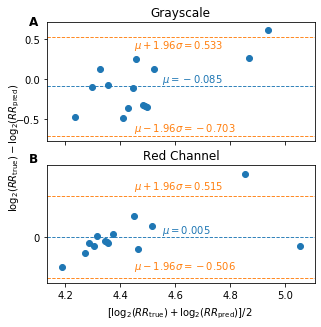}
\caption{Bland-Altman plots for the grayscale (\textbf{A}) and red (\textbf{B}) channeled RR models. Note the relationship between RR and error.}
\label{fig:rrperf-plot}
\end{figure}

For the next set of experiments, the network is tasked with predicting just the RR in breaths per minute. Network performance in this experiment is not significantly different  for both channel types as seen in Table \ref{fig:rrperf-table} and Fig \ref{fig:rrperf-plot}. 

\begin{table}
\begin{center}
\begin{tabular}{c c c c}
Channel & CNN MSE & CNN RMS & EEMD RMS\\
\hline
Gray & 117.30 & $ 10.83$ & $ 19.62$\\
Red & 93.92 & $ 9.69$ & $ 19.35$
\end{tabular}
\end{center}
\caption{DVR 1 Performance (MSE and RMS) of the RR prediction on test data for gray and red channeled models and the EEMD-PCA method. The red channel model performs slightly better than the gray channel, and both improve upon the EEMD-PCA method.}
\label{fig:rrperf1-table}
\end{table}

\begin{table}
\begin{center}
\begin{tabular}{c c c c}
Channel & CNN MSE & CNN RMS & EEMD RMS\\
\hline
Gray & 30.81 & $ 5.56$ & $ 19.62$\\
Red & 23.02 & $ 4.80$ & $ 19.35$
\end{tabular}
\end{center}
\caption{DVR 2 and 3 Performance (MSE and RMS) of the RR prediction on test data for gray and red channeled models and the EEMD-PCA method. The red channel model performs slightly better than the gray channel, and both improve upon the EEMD-PCA method.}
\label{fig:rrperf-table}
\end{table}
For both the red and gray channeled models, the RMS is just under (red) or over (gray) 5 breaths/minute and the Bland-Altman plots suggest that there is reasonable agreement. However, there are clear relationships between the prediction error and the actual RR where lower RRs are overestimated and higher are underestimated. The gray and red channel models achieved correlations of 0.28 and 0.55 respectively between their predictions and the actual test values. 
\newline \indent The discrepancy between the HR and RR model performance on the test set is explained by understanding the task the network must accomplish. In HR prediction, it is fairly straightforward for the 3D filters to learn to estimate the frequency of beats in the video, then transform that frequency to beats per minute. While there are established relationships between features of the PPG and RR (i.e., baseline modulation, amplitude modulation, respiratory sinus arrhythmia), these features a much more subtle than the gross color fluctuations used for determining HR. It is likely that additional noise induced by collecting data during daily life is hiding these features from the network. Nevertheless, the model-based predictions significantly outperform predictions from the EEMD-PCA method on this dataset.     
\subsection{Simultaneous Prediction}
Next, the network is tasked with predicting both the subject HR and RR from a single video clip. As the distribution of HR and RR is not the same, a single loss function cannot be used for this as the distribution of HR and RR is not the same. Therefore, by empirical analysis, the loss function is determined as given below:

\begin{align*}
\mathrm{HRLoss}=\mathrm{(HR(original)-HR(predicted))^{2}} \\
\mathrm{RRLoss}=\mathrm{(RR(original)-RR(predicted))^{2}} \\
\mathrm{TotalLoss}=\mathrm{\sqrt{0.75*HRLoss+0.25*RRLoss}} \\
\end{align*} 

A summary of performance is shown in Table \ref{fig:hrrrperf-table}. For RR, the simultaneous predictor improves performance slightly (RMS of 5.20 vs. 4.80 breaths/minute) when compared to predicting RR alone. However, the model's predictive ability on HR is not as good (RMS of 6.36 vs. 2.92 beats/minute) as the model trained to predict HR alone. This is attributed to the noise added by the respiratory rate task.

\begin{table}
\begin{center}
\begin{tabular}{c c c c c}
Channel & HR MSE & HR RMS & RR MSE & RR RMS \\
\hline
Gray & 199.70 & $ 14.13$ & 96.10 & $ 9.80$ \\
Red & 130.02 & $ 11.40$ & 86.87 & $ 9.32$
\end{tabular}
\end{center}
\caption{DVR 1 Performance (MSE and RMS) of the simultaneous HR and RR predictions on the test data for the two-channel types.}
\label{fig:hrrrperf1-table}
\end{table}

 \begin{table}
\begin{center}
\begin{tabular}{c c c c c}
Channel & HR MSE & HR RMS & RR MSE & RR RMS \\
\hline
Gray & 93.70 & $ 9.68$ & 27.02 & $ 5.20$ \\
Red & 40.45 & $ 6.36$ & 37.65 & $ 6.14$
\end{tabular}
\end{center}
\caption{DVR 2/3 Performance (MSE and RMS) of the simultaneous HR and RR predictions on the test data for the two channel types.}
\label{fig:hrrrperf-table}
\end{table}

\indent Given the performance shown in Table. \ref{fig:hrrrperf-table}, the top-performing red channel model for each subtask in simultaneous prediction is analyzed via Bland-Altman. This analysis shows that this approach is able to reduce the relationship between prediction error and the actual RR. This is explained by the extra information encoded within two outputs rather than one: the HR prediction informs the RR prediction and vice versa.

\begin{figure}
\includegraphics[width=0.8\textwidth]{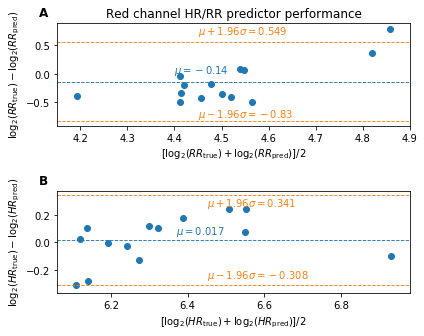}
\caption{Bland-Altman plots of the (top performing) red channel simultaneous predictor on RR (\textbf{A}) and for HR (\textbf{B}). Note that this model seems to reduce the relationship between prediction error and RR.}
\label{fig:hrrr_perf-plot}
\end{figure}

\section{Discussion}

\indent Herein, a novel dataset comprised of mobile phone video recordings captured during daily life, annotated with ground truth HR and RR labels has been introduced. By employing a recent signal-processing-based method (EEMD-PCA) – previously applied only to data gathered under controlled conditions – this study finds that this approach can estimate HR with an RMS error of 9 beats/minute (see Table \ref{fig:hrperf-table}) and RR with an RMS error of 19 breaths/minute (see Table \ref{fig:rrperf-table}). These findings encourage the development of an alternative, more robust method for estimating HR and RR based on deep video regression. The proposed new method surpasses the EEMD-PCA approach, achieving a 68\% reduction in HR RMS (3 beats/minute, see Table \ref{fig:hrperf-table}) and a 75\% reduction in RR RMS (5 breaths/minute, see Table \ref{fig:hrperf-table}) on this dataset. Furthermore, this research delves into various network configurations for estimating HR and RR and evaluate their performance using Bland-Altman and correlation plots.
\newline \indent The error rates for the EEMD-PCA method presented here are considerably higher than those documented in the literature (maximum values from \cite{s3}: RR RMS of 2.7 vs. 19 breaths/minute, HR RMS of 0.69 vs. 9 beats/minute). This indicates that the method's performance may not generalize well to noisy, real-world PPG analogs derived from mobile phone videos. However, deep video regression substantially improves these results.The accuracy of the HR predictions is nearing the expected error in the gold standard labels (a mean absolute error of just over 4 beats/minute, as per \cite{b20}). Furthermore, Bland-Altman and correlation analyses reveal excellent agreement across the range of observed HRs. While the RR prediction performance surpasses the EEMD-PCA method, it does not perform as well as HR overall. Future research should contemplate utilizing longer window lengths (30 seconds) to determine if the network better resolves the RR-related features of the PPG (i.e., baseline modulation, amplitude modulation, respiratory sinus arrhythmia). Nonetheless, the relative success of the proposed approach implies that deep video regression merits further investigation for estimating HR and RR from noisy, real-world mobile phone videos.
\newline \indent Interpreting these findings in the context of existing literature is crucial. Studies have indicated the feasibility of extracting HR and RR from mobile phone fingertip videos within a margin of several beats or breaths per minute error. However, data collection in these studies typically occurs in laboratory environments under strict experimental control, despite the ultimate aim of deploying these methods in daily life. Many assume these methods will generalize, but the present results challenge this notion. As mentioned in the Methods section, 56\% of videos were discarded due to the absence of a distinguishable "pulsing" or "beating" pattern. After an informal examination of the discarded videos, interviews with several subjects, and observations of multiple collections, the poor quality of discarded videos can be attributed to 1) inattention to instruction, 2) incorrect finger placement on the camera lens, and 3) applying excessive or insufficient pressure. Moreover, even for the remaining high-quality videos, standard algorithms demonstrated subpar performance on this dataset, exhibiting significantly higher errors in HR and RR predictions compared to previously published studies.
\newline \indent As research progresses in the realm of deep video regression for estimating HR and RR, future studies should contemplate using longer video clips to more accurately resolve the RR-related photoplethysmography (PPG) features. Furthermore, to enhance data quality, it may be advisable to create mobile phone accessories designed to assist users in accurately positioning their fingers on the camera lens and applying an appropriate level of pressure when measuring vital signs.
\section{Conclusion}
\indent This study presents a new dataset of mobile phone video recordings made during daily life and annotated with ground truth HR and RR labels. The poor performance of an existing algorithm for estimating HR and RR from these data motivates the development of a new method that employs deep video regression. Results demonstrate that this method improves HR prediction performance by 68\% and RR prediction performance by 75\%. Future work should examine the effect of video length on these results.

\section{Acknowledgements}
The authors would like to thank dtectron2 (\url{https://github.com/facebookresearch/detectron2}) and Daniel Berenberg (https://github.com/djberenberg/wepanic) for giving access to the pre-trained models, supplementary scripts which helped us to make the data processing work more feasible.

\newpage
\renewcommand{\refname}{\centering \normalsize REFERENCES}

\bibliographystyle{plainnat}
\bibliography{trb_template}

\end{document}